\documentclass{egpubl}
\usepackage{eurovis2020}

\EuroVisShort  % for EuroVis additional material

\usepackage[T1]{fontenc}
\usepackage{dfadobe}  

\usepackage{cite}  % comment out for biblatex with backend=biber
\BibtexOrBiblatex
\electronicVersion
\PrintedOrElectronic
\ifpdf \usepackage[pdftex]{graphicx} \pdfcompresslevel=9
\else \usepackage[dvips]{graphicx} \fi

\usepackage{egweblnk}
\usepackage{subfigure} 
\usepackage{amsmath,amssymb}
\usepackage[table]{xcolor}
\usepackage{moresize}
\usepackage[utf8]{inputenc}

\newcommand{\Rspace}        {{\mathbb R}}

\newcommand {\mm}[1] {\ifmmode{#1}\else{\mbox{\(#1\)}}\fi}

\newcommand{\Fgroup}        {{\mathcal{F}}}

\newcommand{\BigO} {\mathcal{O}}

\usepackage{tikz}
\newcommand*\circled[1]{\tikz[baseline=(char.base)]{
            \node[shape=circle,draw,inner sep=0.5pt] (char) {#1};}}

\newcommand{\circledSymbol}[1]{{\fontsize{6}{7}\circled{#1}}}

\title[TopoLines: Topological Smoothing for Line Charts]%
      {TopoLines: Topological Smoothing for Line Charts}

\author[Rosen et al.]
{\parbox{\textwidth}{\centering 
            P.\ Rosen$^{1}$\orcid{0000-0002-0873-9518}
        and A.\ Suh$^{1,2}$\orcid{0000-0001-6513-8447}
        and C.\ Salgado$^{1}$
        and M.\ Hajij$^{3}$\orcid{0000-0002-2625-9286}
        }
        \\
{\parbox{\textwidth}{\centering 
         $^1$University of South Florida, Tampa FL, USA \\
         $^2$Tufts University, Medford, MA, USA \\
         $^3$KLA Tencor, Ann Arbor, MI, USA
       } 
}
}

\begin{document}

\teaser{
    \centering
    
    \begin{minipage}[c]{5pt}
        \rotatebox{90}{\scriptsize \hspace{5pt} High Smoothing \hspace{10pt} Low Smoothing \hspace{20pt} Original}
    \end{minipage}
    \begin{minipage}[c]{5pt}
        \rotatebox{90}{\scriptsize \hspace{5pt} (80\%) \hspace{35pt} (40\%) \hspace{40pt}}
    \end{minipage}  
    \hspace{-5pt}
    \begin{minipage}[c]{0.86\linewidth}
        \includegraphics[width=1\linewidth]{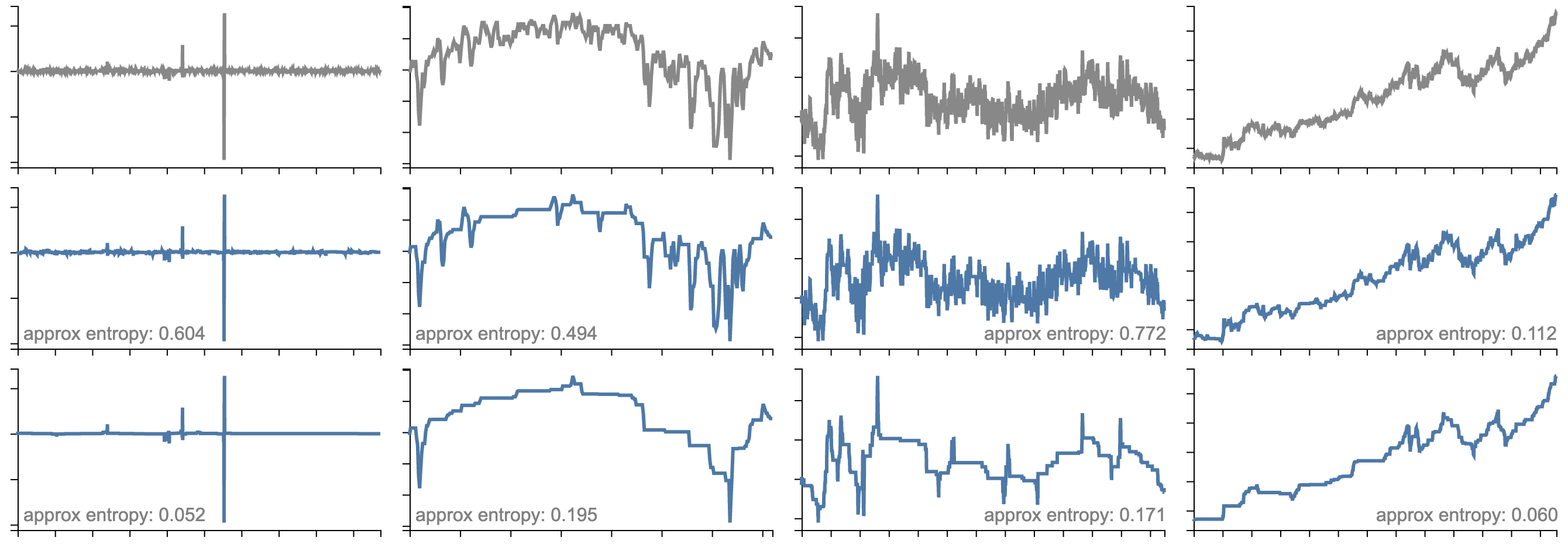}
        
        \vspace{-15pt}
        \subfigure[Astro 115/128\label{fig:datasets:astro}]{%
            \begin{minipage}[c]{0.235\linewidth} \ \end{minipage}%
        }
        \hfill
        \subfigure[Climate 17/18\label{fig:datasets:clim}]{%
            \begin{minipage}[c]{0.235\linewidth} \ \end{minipage}%
        }
        \hfill
        \subfigure[EEG Chan 7\label{fig:datasets:eeg}]{%
            \begin{minipage}[c]{0.235\linewidth} \ \end{minipage}%
        }
        \hfill
        \subfigure[Stock GOOG\label{fig:datasets:goog}]{%
            \begin{minipage}[c]{0.235\linewidth} \ \end{minipage}%
        }
    \end{minipage}

    %\vspace{-4pt}
    \caption{Examples of the (top) 4 input datasets (see \autoref{sec:results} for a description of the data). TopoLines results are shown for (middle) low and (bottom) high levels of smoothing, defined by the percent of local extrema removed. TopoLines works by preserving high amplitude extrema and flattening low amplitude ones while maintaining low residual error. See the supplement for the measures described in \autoref{sec:eval_methodology}.}
    \label{fig:datasets}
}

\maketitle

\begin{abstract}
   Line charts are commonly used to visualize a series of data values. When the data are noisy, smoothing is applied to make the signal more apparent. Conventional methods used to smooth line charts, e.g., using subsampling or filters, such as median, Gaussian, or low-pass, each optimize for different properties of the data. The properties generally do not include retaining peaks (i.e., local minima and maxima) in the data, which is an important feature for certain visual analytics tasks. We present TopoLines, a method for smoothing line charts using techniques from Topological Data Analysis. The design goal of TopoLines is to maintain prominent peaks in the data while minimizing any residual error. We evaluate TopoLines for 2 visual analytics tasks by comparing to 5 popular line smoothing methods with data from 4 application domains. \\
%   Leave one blank line after the abstract, 
%   then add the subject categories according to the ACM Classification Index 
%-------------------------------------------------------------------------
%  ACM CCS 1998
%  (see https://www.acm.org/publications/computing-classification-system/1998)
% \begin{classification} % according to https://www.acm.org/publications/computing-classification-system/1998
% \CCScat{Computer Graphics}{I.3.3}{Picture/Image Generation}{Line and curve generation}
% \end{classification}
%-------------------------------------------------------------------------
%  ACM CCS 2012 (see https://www.acm.org/publications/class-2012)
%The tool at \url{http://dl.acm.org/ccs.cfm} can be used to generate
% CCS codes.
%Example:



\end{abstract}

%\keywords{Line chart, data smoothing, Topological Data Analysis.}

\section{Introduction}

Line charts are used to analyze data in a variety of applications, including identifying stock trends, tracking weather changes, understanding brain activity, etc. While significant increases in data availability allow users to create plots with many data points, relieving visual clutter requires performing additional data processing, such as smoothing. However, the way smoothing modifies the data can have an impact on the performance of visual analytics tasks. We consider smoothing in the context of 2 low-level tasks~\cite{amar2005low}, finding extrema (i.e., local minima and maxima) and retrieving a value. These tasks, in essence, require that any smoothing method both retain extrema and minimize any residual error they introduce (i.e., the difference between the input and output data).

\begin{figure*}[!t]
    \begin{minipage}[b]{0.245\linewidth}
        \hfill
        \begin{minipage}[t]{0.0\linewidth}
            \begin{minipage}[t]{375pt} 
        
                \vspace{-30pt}
                \subfigure[\label{fig:ph_ex:input}]{}
                
                \vspace{10pt}
                \subfigure[\label{fig:ph_ex:output}]{}

            \end{minipage}
        \end{minipage}    
        \begin{minipage}[t]{0.785\linewidth}
            \centering
            \includegraphics[width=0.725\linewidth]{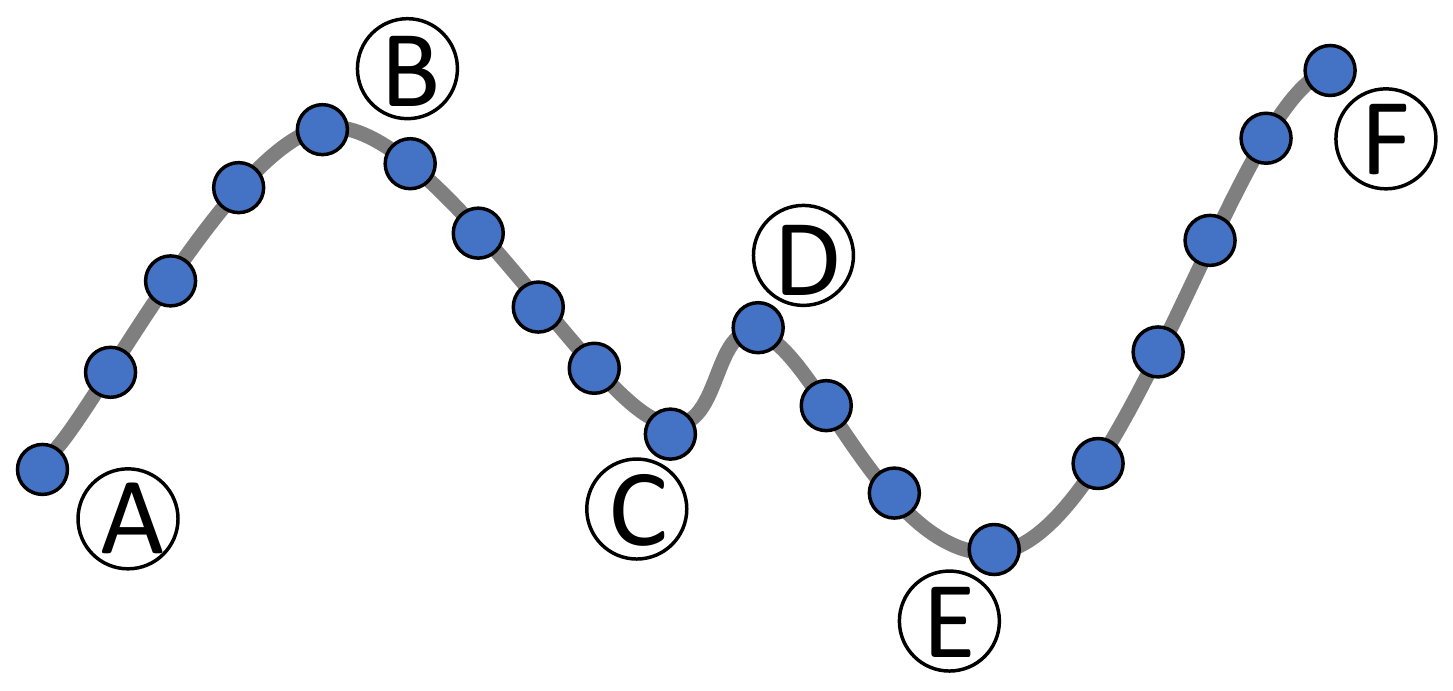}
            
            \includegraphics[width=0.725\linewidth]{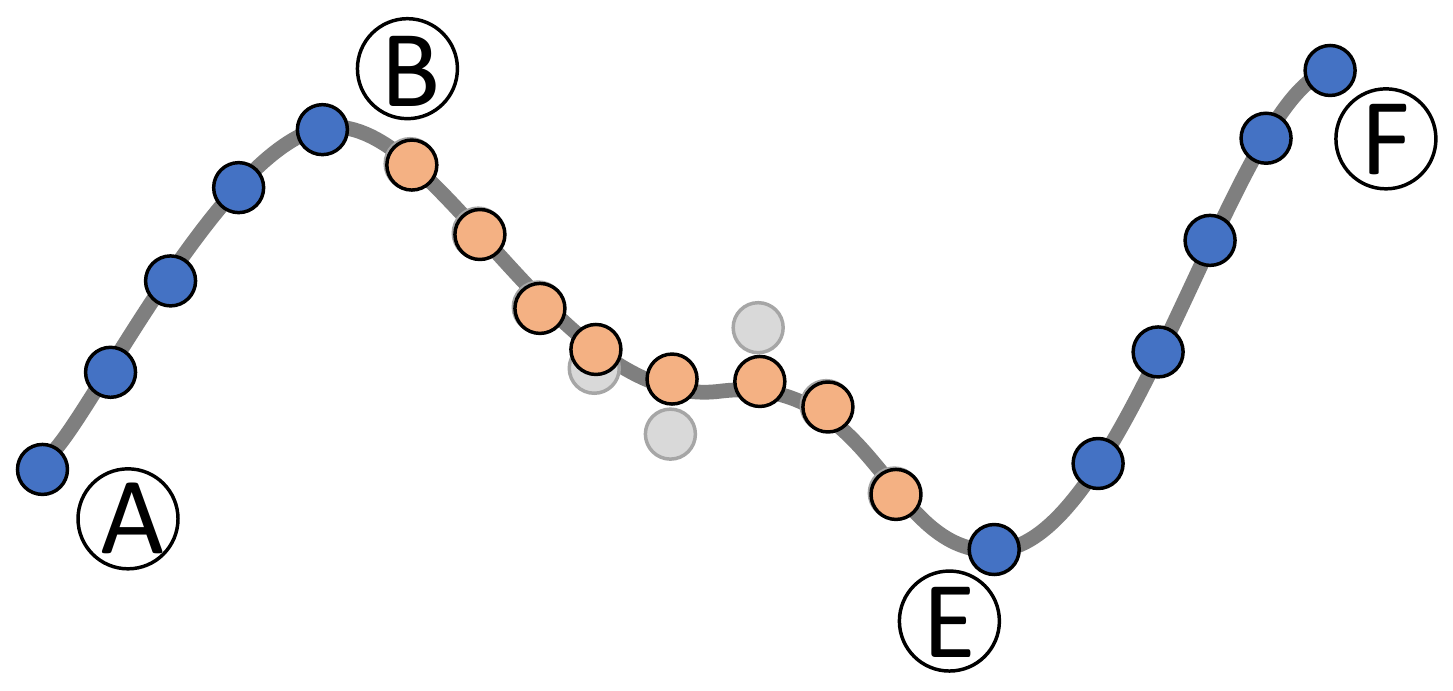}
        \end{minipage}
        \hspace{5pt}
        \caption{(a) Input line chart (b)~after removing an extrema pair.}
        \label{fig:smoothed_ex}    
    \end{minipage}
    \hfill
    \begin{minipage}[b]{0.725\linewidth}
        \centering
        {
        \begin{minipage}[c]{1\linewidth}
            %\hspace{15pt}
            \begin{minipage}[t]{0.985\linewidth}
                \includegraphics[width=0.31\linewidth]{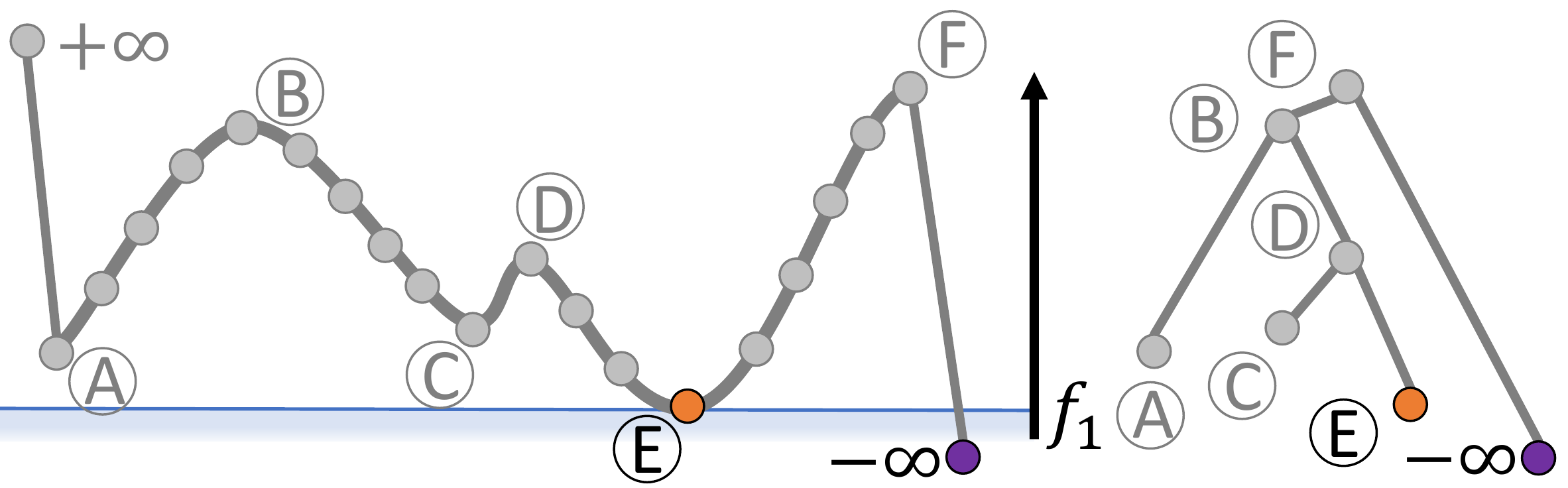}
                \hfill
                \includegraphics[width=0.31\linewidth]{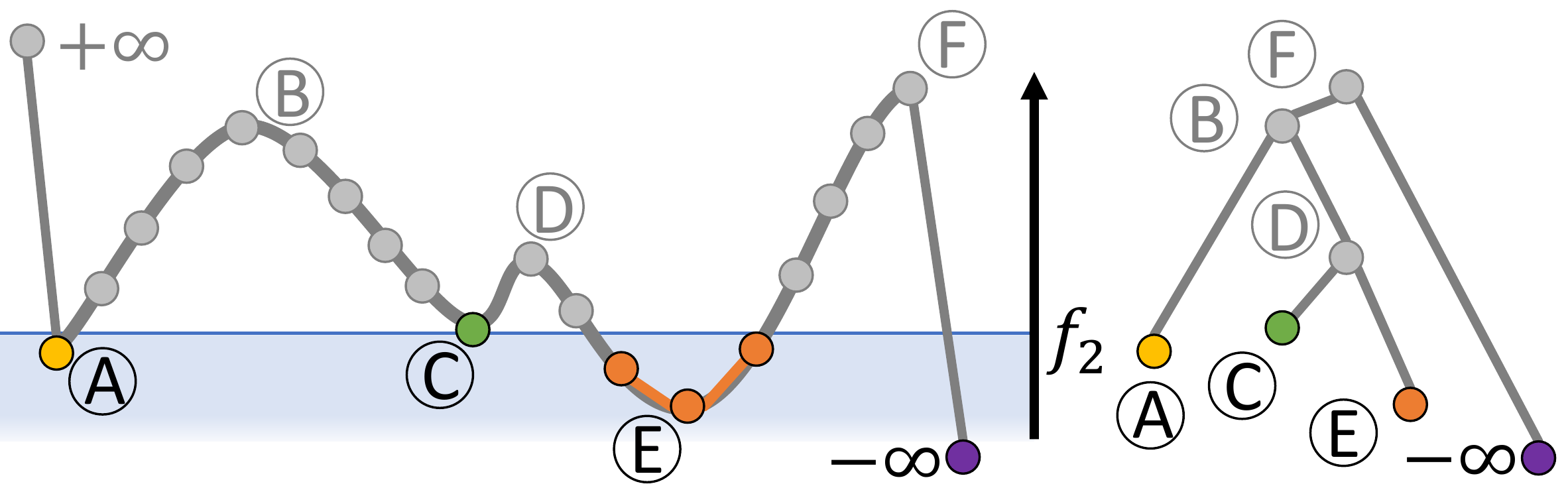}
                \hfill
                \includegraphics[width=0.31\linewidth]{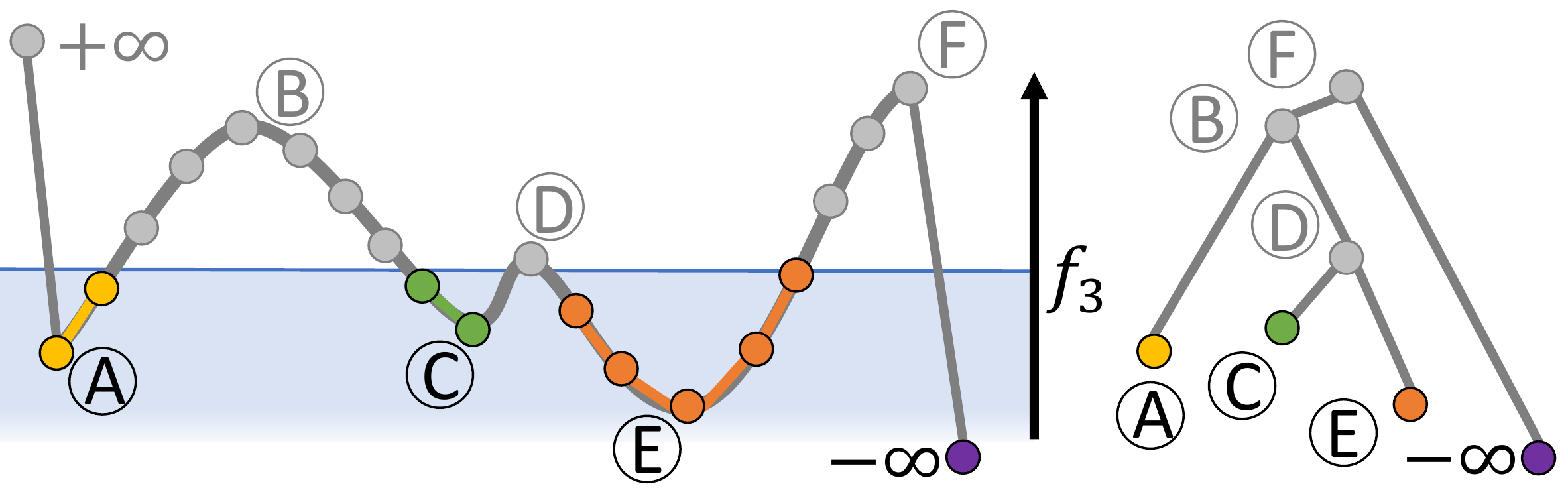}
                
                \vspace{5pt}
                \includegraphics[width=0.31\linewidth]{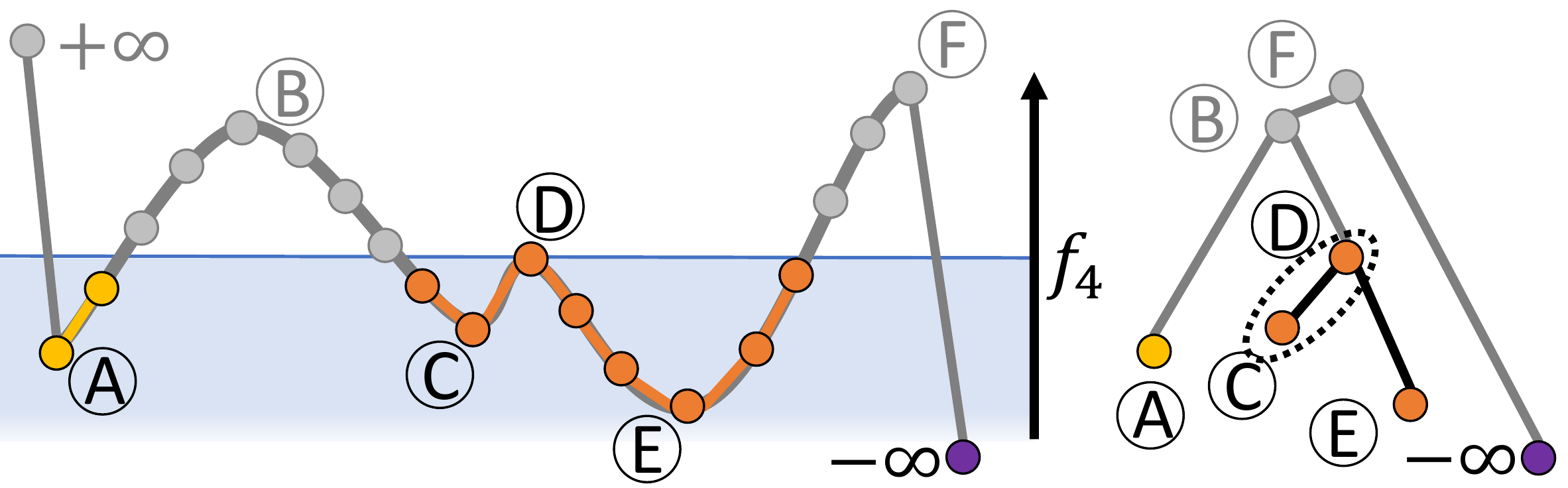}
                \hfill
                \includegraphics[width=0.31\linewidth]{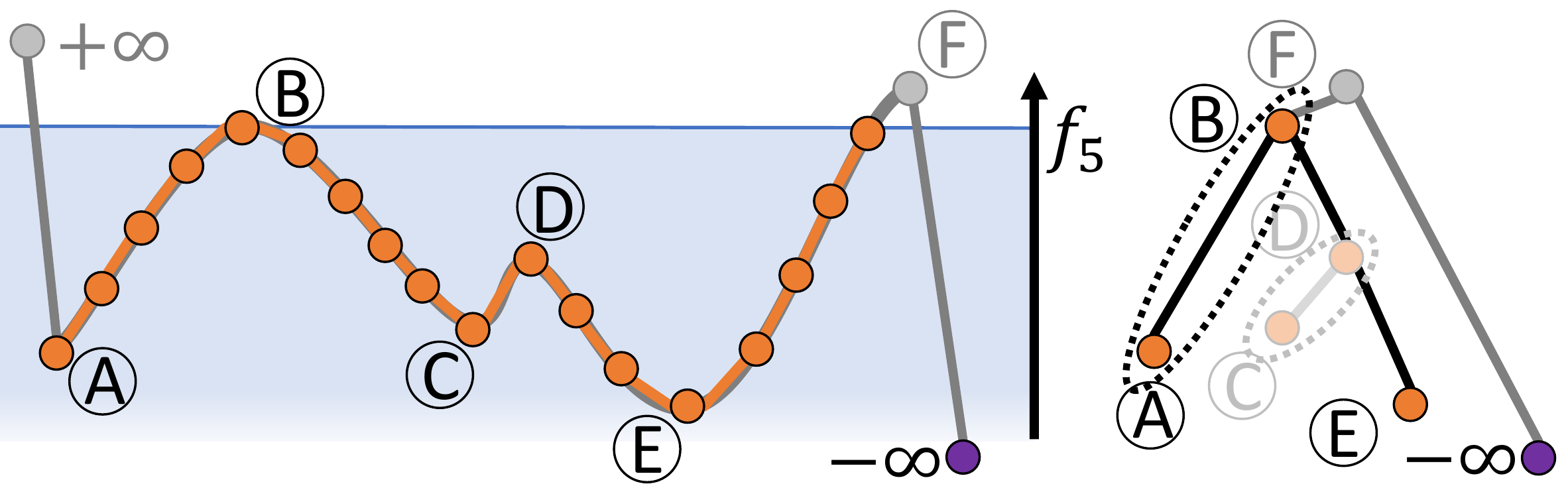}
                \hfill
                \includegraphics[width=0.31\linewidth]{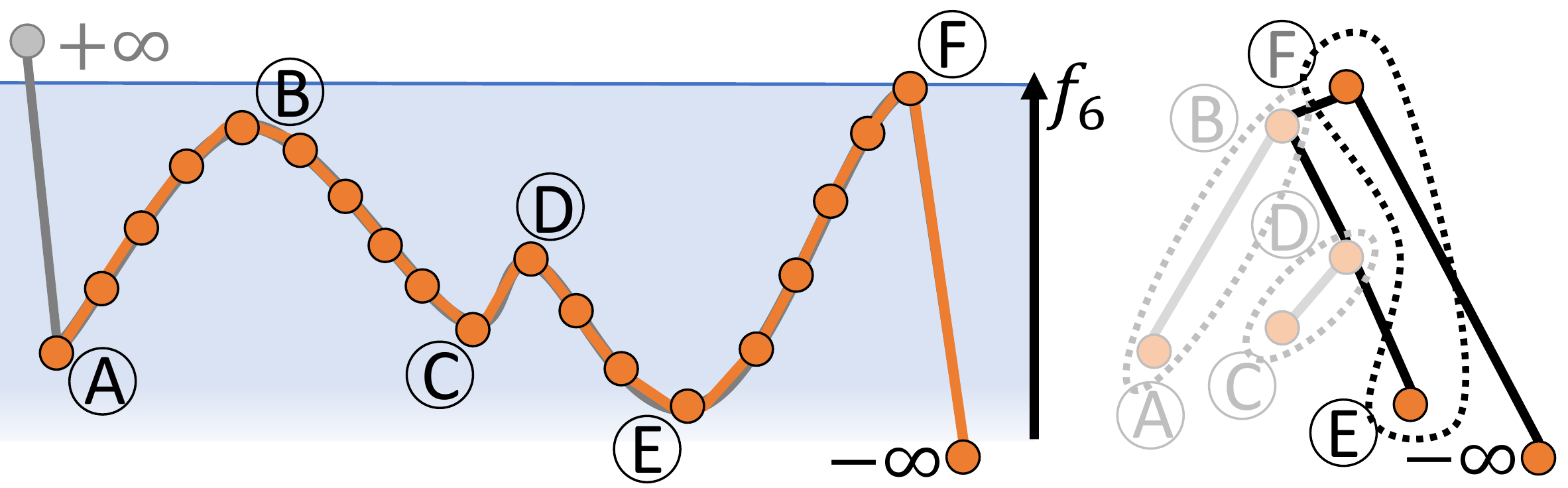}        
                
            \end{minipage}
            \hspace{-375pt}
            \begin{minipage}[t]{0.0\linewidth}
                \begin{minipage}[t]{400pt} 
            
                    \vspace{-45pt}
                    \subfigure[\label{fig:ph_ex:a1}]{}
                    \hspace{120pt}
                    \subfigure[\label{fig:ph_ex:a2}]{}
                    \hspace{120pt}
                    \subfigure[\label{fig:ph_ex:a3}]{}
                
                    \vspace{9pt}
                    \subfigure[\label{fig:ph_ex:a4}]{}
                    \hspace{120pt}
                    \subfigure[\label{fig:ph_ex:a5}]{}
                    \hspace{120pt}
                    \subfigure[\label{fig:ph_ex:a6}]{}
    
                \end{minipage}
            \end{minipage}
        \end{minipage}
        }
        \caption{Six steps of a lower-star filtration (left) and merge tree construction (right) on the line chart.}
        \label{fig:ph_ex}
    \end{minipage}
\end{figure*}

\begin{comment}
\begin{figure}[!b]

        \hspace{8pt}
        \begin{minipage}[t]{0.925\linewidth}
            \hfill
            \includegraphics[width=0.475\linewidth]{figures/topoline_1.pdf}
            \hspace{5pt}
            \includegraphics[width=0.475\linewidth]{figures/topoline_2.pdf}
        \end{minipage}
        \hspace{-223pt}
        \begin{minipage}[t]{0.0\linewidth}
            \begin{minipage}[t]{400pt} 
        
                \vspace{-60pt}
                \subfigure[\label{fig:ph_ex:input}]{}
                \hspace{115pt}
                \subfigure[\label{fig:ph_ex:output}]{}

            \end{minipage}
        \end{minipage}    
    \caption{(a) Input line chart after (b) removing an extrema pair.}
    \label{fig:smoothed_ex}
\end{figure}
\end{comment}

A variety of smoothing techniques are available. Uniform subsampling, for example, skips data on a regular interval, and while trivial to implement, the output optimizes upon no particular quality of the input. Other common methods, such as median, Gaussian, and low-pass cutoff filters, retain low-frequency aspects of the data but potentially lose extrema in the data. Irregular sampling, such as Douglas-Peucker~\cite{ramer1972iterative,douglas1973algorithms}, does a better job preserving extrema, but it retains little detail in the smoothing process.

We address the weaknesses of prior approaches by applying Topological Data Analysis (TDA) to line chart smoothing. We do this by using TDA to capture a hierarchical relationship between extrema that allows removing those of ``low importance''. At the same time, TopoLines minimizes the residual error between extrema, retaining much of the detail from the input data.

Previously, Kozlov and Weinkauf released \textit{Persistence1D}, a TDA-based class for filtering 1D data using their persistence~\cite{yeara}. There has also been work done regarding topological smoothing of 2D and 3D functions~\cite{CarrSnoeyinkPanne2010,EdelsbrunnerMorozovPascucci2006,rosen2019using,ttk}. However, we could find no prior studies that compared topological smoothing to conventional techniques in line charts. Therefore, our contributions are: 1) a description of 1D topological smoothing; 2) optimizations of topological smoothing for the visual analytics tasks of retrieving a value and finding extrema; and 3) an analytical evaluation of the effectiveness of TopoLines and 5 conventional smoothing methods on 4 dataset types.

Our results show that TopoLines is the most effective approach for many, but not all, combinations of data type and task. Almost as important, our results demonstrate the general ineffectiveness of several conventional methods, including median filters, cutoff filters, and uniform subsampling in the tasks and data evaluated.

\section{TopoLines: Topologically Smoothed Line Charts}

TopoLines smoothing requires 2 steps: 1)~extraction of the topology of the data using persistent homology, and 2)~smoothing the output by removing extrema based upon a user-selectable threshold.

\subsection{Persistent Homology of a Line Chart}
\label{sec:topolines:ph}

We provide a practical description of persistent homology using the line chart in \autoref{fig:smoothed_ex} and \ref{fig:ph_ex} as an example while leaving further details and theoretical justifications to~\cite{EdelsbrunnerHarer2008}.

We use the lower-star filtration of the simplicial complex, $\Fgroup$, i.e., the points and edges, on the function $f:\Fgroup \rightarrow \Rspace$. The lower-star filtration of the data tracks the creation and merging of connected components of the sublevelset $|\Fgroup|_i=f^{-1}(-\infty,f_i]$, as $f_i$ is swept from $-\infty \rightarrow \infty$, represented by the blue region in \autoref{fig:ph_ex}. The filtration is calculated by first sorting the points of $f$ in increasing order. Then, points are inserted into $|\Fgroup|$ one at a time. An edge is added between any neighboring points already in $|\Fgroup|_i$.

The relationship between connected components is tracked using a merge tree parameterized by $f$. When a component first appears at $f_i$, caused by a local minimum, a leaf node is added to the merge tree at $f_i$. For example in \autoref{fig:ph_ex:a1}, the orange connected component is formed at~\circledSymbol{E}, and an equivalent leaf node is created in the merge tree. As the plane is swept higher, as in \autoref{fig:ph_ex:a2}, new connected components---\circledSymbol{A} in yellow and \circledSymbol{C} in green---are created.

\begin{figure*}[!t]
    \includegraphics[width=1\linewidth]{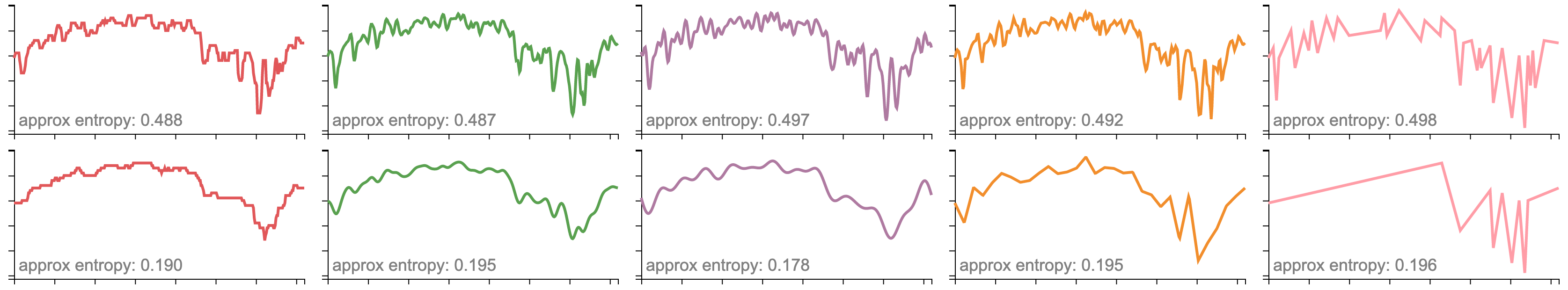}
    
    \vspace{-15pt}
    \noindent
    \subfigure[Median\label{fig:otherFilters:med}]{\hspace{0.197\linewidth}}
    \subfigure[Gaussian\label{fig:otherFilters:gau}]{\hspace{0.197\linewidth}}
    \subfigure[Cutoff\label{fig:otherFilters:lp}]{\hspace{0.197\linewidth}}
    \subfigure[Uniform Subsampling\label{fig:otherFilters:subs}]{\hspace{0.197\linewidth}}
    \subfigure[Douglas-Peucker\label{fig:otherFilters:dp}]{\hspace{0.197\linewidth}}

    \caption{Conventional smoothing on Climate 17/18 with approximate entropy similar to  \autoref{fig:datasets:clim}(middle and bottom).}
    \label{fig:otherFilters}
\end{figure*}

When 2 components merge, representing a local maximum, a merge node is created in the merge tree at $f_i$ and connected to the merged components. In \autoref{fig:ph_ex:a3}/\ref{fig:ph_ex:a4}, the green and orange connected components merge at \circledSymbol{D}, a local maximum. The connected components are combined, in orange, and a merge node is added to the merge tree. 
When a merge node is created, it is also paired with a leaf node (i.e., a local maximum is paired with a local minimum). In particular, it is paired with the minimum from the two merging components with the \textit{larger} value. Referring to \autoref{fig:ph_ex:a3}/\ref{fig:ph_ex:a4}, the point \circledSymbol{D} is paired with the minimum from the green and orange components with the larger value, in this case point \circledSymbol{C}. In other words, $f(\circledSymbol{C})>f(\circledSymbol{E})$, therefore, $[\circledSymbol{C},\circledSymbol{D})$ form an extrema pair. The new merged component in orange continues with minimum \circledSymbol{E}. Similarly, in \autoref{fig:ph_ex:a4}/\ref{fig:ph_ex:a5}, at \circledSymbol{B}, the value of the minimum of yellow $f(\circledSymbol{A})$ and orange $f(\circledSymbol{E})$ are compared, and $[\circledSymbol{A},\circledSymbol{B})$ are paired. 
The output of the operation is the set of all extrema pairs, $C=\{[b_0,d_0), [b_1,d_1),...,[b_m,d_m)\}$, where $b_i$ and $d_i$ are the local minimum and maximum, respectively, and $m$ is the number of pairs. 

Boundaries require special handling, as notable in \autoref{fig:ph_ex}. If a boundary point is a local minimum, e.g., $\circledSymbol{A}$, it is connected to a point at $+\infty$. Similarly, a local maximum boundary point is connected to $-\infty$, e.g., $\circledSymbol{F}$. The additional points ensure all extrema are paired. The algorithm has $\BigO(n \log{} n)$ complexity by using the disjoint-set data structure to track connected components. The complexity improves to $\BigO( n + m \log{} m)$ by removing all non-extrema from the input before merge tree construction.

\subsection{Topological Simplification}

The set of extrema pairs, $C$, is used to guide smoothing, as follows. For each pair, a measure known as \textit{persistence} is calculated, which is simply the difference in function value between the local minimum and local maximum of the pair, i.e., $p_i=|f(d_i)-f(b_i)|$. In effect, this measures the \textit{peak-to-peak amplitude}. 

The simplification is controlled by removing extrema pairs from the output through either a user-specified persistence threshold, $t$, to remove pairs, $\{C_i | p_i<t\}$, or by removing a percentage, $q$, of pairs by ranking/sorting them, $\{C_i | rank(C_i)<q \cdot m\}$. To reconstruct the line, the extrema that are not removed, in addition to the boundary points, are first placed into the output. For \autoref{fig:ph_ex:output}, this includes $\circledSymbol{A}$, $\circledSymbol{B}$, $\circledSymbol{E}$, and $\circledSymbol{F}$. Next, the intermediate data is calculated. 

As pointed out by prior work on 2D manifolds~\cite{EdelsbrunnerMorozovPascucci2006} and contour trees~\cite{CarrSnoeyinkPanne2010}, removing a pair of critical points from the function is as simple as ``flattening'' the function. 
For a 1D function, this equates to making the function monotonic between neighboring extrema. For example, in \autoref{fig:ph_ex:output}, removing the $[\circledSymbol{C},\circledSymbol{D})$ critical point pair requires modifying the function such that it is monotonically decreasing between critical points \circledSymbol{B} and \circledSymbol{E}.

The design space of possible modifications is quite broad---\textit{any} monotonic function satisfies the topological constraint. We apply the additional constraint that the remainder of the function is modified as little as possible. To accomplish this, we use isotonic regression~\cite{barlow1972statistical}, which is a monotonic regression technique that minimizes the least square error. The time complexity of isotonic regression and our reconstruction is $\BigO(n)$.

\begin{figure*}[!th]
    \centering
    \includegraphics[width=0.975\linewidth]{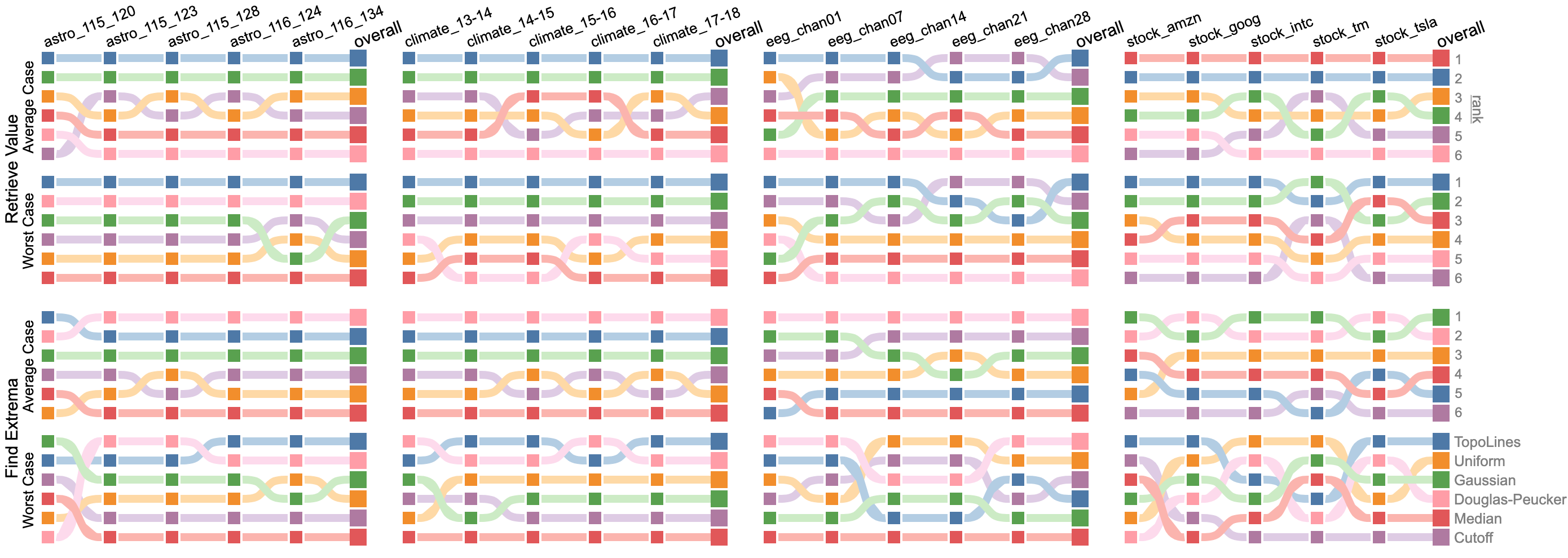}
    \vspace{-3pt}
    \caption{Ranking of all methods for all data and both tasks. Overall rank is determined by the average rank for the connected datasets/tasks.}
    \label{fig:ranks}
    \vspace{-4pt}
\end{figure*}

\section{Evaluation}
\label{sec:eval_methodology}

We compare TopoLines to 5 other smoothing methods: 

\begin{itemize} 

\item A \textsc{median filter} (see \autoref{fig:otherFilters:med}) is a nonlinear rank filter, which is particularly good at removing salt-and-pepper noise~\cite{arce2005nonlinear}. For each input datum, the filter extracts a surrounding neighborhood window and outputs its median value. Smoothing is increased by enlarging the window size.

\item The \textsc{Gaussian filter} (see \autoref{fig:otherFilters:gau}) a commonly used  {convolutional filter} in signal and image processing~\cite{gaussian}.
The approach applies a stencil, whose weights come from a normal distribution, to an input neighborhood.  The smoothing level is changed by adjusting the standard deviation of the distribution.

\item A low-pass \textsc{cutoff filter} (see \autoref{fig:otherFilters:lp}) converts the scalar data into the frequency domain via Discrete Fourier Transform (DFT)~\cite{cooley1965algorithm}, zeros frequencies above a cutoff threshold, and computes the new scalar values with an inverse DFT.  The level of smoothing is adjusted by modifying the cutoff frequency.

\item \textsc{Uniform subsampling} (see \autoref{fig:otherFilters:subs}) selects points at regular intervals. Between selected points, linear interpolation is used. The smoothing level is increased by sampling fewer points.

\item \textsc{Douglas-Peucker}~\cite{ramer1972iterative,douglas1973algorithms} (see \autoref{fig:otherFilters:dp}) is a non-uniform subsampling approach that optimizes the $l^\infty$-norm of the residual error. The algorithm starts by selecting the boundary points of the input and connects them with linear interpolation. Points are then iteratively added by inserting the input point with the largest distance to the output. The process repeats until a user-specified threshold distance is reached.

\end{itemize}

\subsection{Task Analysis}
\label{sec:eval_methodology_tasks}

We considered a variety of low-level tasks based upon the taxonomy of Amar et al.~\cite{amar2005low} and settled upon 2 tasks that we hypothesized TopoLines would perform well. For each, we only consider the resulting impact on the modification of the data, not the perceptual impact of the smoothing (see future work in \autoref{sec:conclusions}). For each task, we provide a brief description along with average and worst case analytical measures of performance.

\paragraph*{Retrieve Value} is a task focused on finding a specific function value on a given chart. An example query would be, ``What was the GOOG stock (\autoref{fig:datasets:goog}) price on April 15, 2018?'' The accuracy of retrieving a value is dependent upon how closely the values of the smoothed data reflect the values in the input data. We measure this by considering the residual error between the original and smoothed data using vector norms. 

For the \textbf{average case} performance, we consider the $l^1$-norm: $\left\|\mathbf {l} \right\|_{1}=\sum _{i=1}^{n}\left|x_{i}-x'_{i}\right|$, which measures the sum of the absolute value of errors. Since the data length is fixed, comparing the sum of errors is equivalent to comparing the average error. 
For the \textbf{worst case} performance, we consider the $l^\infty$-norm: $\left\|\mathbf {l} \right\|_{\infty}=\max\limits_i \left|x_{i}-x'_{i}\right|$, which measures only the point of the largest difference between the input and output data.

\paragraph*{Find Extrema} task is concerned with identifying minima and maxima in the data. An example query would be, ``What are the dates of the top 3 peaks of GOOG (\autoref{fig:datasets:goog})?'' The performance of this task requires that in smoothing, extrema remain in the data. To measure the performance, we calculated the topological difference between the input and smoothed data using methods from TDA~\cite{EdelsbrunnerHarer2010}.
First, the persistent homology of the original and smoothed data are calculated, as described in \autoref{sec:topolines:ph}, to create 2 sets of extrema pairs $C$ and $C'$, respectively. For technical reasons, all pairs with infinite persistence are removed, and all pairs of 0-persistence $[c,c)$ are added to make the cardinality infinite~\cite{kerber2017geometry}. Let $\eta$ be a bijection between the 2 sets. 

The \textbf{average case} is measured using the 1-Wasserstein distance, 
$W_1(C,C') = \displaystyle \inf_{\eta:C \rightarrow C'}  \Sigma_{c\in C} \left\lVert c-\eta(c) \right\rVert_1$, between the input and output extrema pairs, which identifies the average perturbation of extrema. The \textbf{worst case} is measured using Bottleneck distance, $W_{\infty}(C,C') = \inf_{\eta: C \rightarrow C'} \sup_{c \in C} \left\lVert c-\eta(c) \right\rVert_\infty$, which only returns the difference in the extrema with the largest distortion.

\paragraph*{Baseline.} Each smoothing method offers an adjustable simplification parameter, whose interpretation and output are approach dependent. This variation prevents us from directly using the threshold for comparing methods.
Instead, we use approximate entropy as a calibration measure since it has been shown to be a good proxy for line chart complexity~\cite{2018entropy} (see \autoref{fig:otherFilters}).

\paragraph*{Comparison.} To compare methods, we evaluated each technique using the 4 metrics, described above, across the full range of approximate entropy values. Each technique/metric then had the best fit line calculated, and the approaches were ranked by their area under the curve from smallest to largest. In other words, for a given measure, the methods are ranked by which produces the lowest error across the range of entropy values. See the supplemental materials for all measures and best fit lines.

\section{Results and Discussion}
\label{sec:results}

We test our method using 4 application domains (see \autoref{fig:datasets}) of 5 datasets each.
%: radio \textit{astro}nomy, \textit{climate}, electroencephalogram (\textit{EEG}), and \textit{stock} trends. For each domain, we tested 4 datasets. 
%
Radio \textit{astro}nomy data are 5 spectral ``lines'' that measure the frequency and amplitude of radio waves emitted by extraterrestrial matter (i.e., gas and dust) and was downloaded from~\cite{almaData}.
\textit{Climate} is a measure of daily high temperature recorded from July to July over 5 periods (20-13/14 through 20-17/18) at a large metropolitan airport downloaded from~\cite{young_knapp_inamdar_hankins_rossow_2018}. 
The \textit{EEG} data each contain a window from 5 (of 32 total) channels of a single subject undergoing a visual attention task and was acquired from~\cite{eegData}.
\textit{Stock} trends contain daily closing values for 5 companies (Amazon, Google, Intel, Toyota, and Tesla) over 5 years, starting in February 2015, collected from Yahoo Finance.
%
%Examples for each domain, along with TopoLines smoothing, can be found in \autoref{fig:datasets}. 
%
%Source code is available at \textit{$<$withheld for double-blind review$>$}.
All source code is available at \url{https://github.com/USFDataVisualization/TopoLines}, and results are available at \url{https://USFDataVisualization.github.io/TopoLines}. 

%Our study was run on Amazon's Mechanical Turk and consisted of 36 participants. All participants performed both tasks outlined in \autoref{sec:eval_methodology_tasks} across all datasets and smoothing methods. A detailed summary of our user study is available in the supplementary materials. 
The results for all data, measures, and smoothing approaches are summarized in \autoref{fig:ranks}. 
For all datasets, TopoLines performed best in both average and worst case for the retrieve value task, with the only exception being a second-best finish for average case with the \textit{stock} data. For the find extrema task, TopoLines performed best or second-best in the average and worst cases for \textit{astro} and \textit{climate} data. For the \textit{EEG} and \textit{stock} data, TopoLines performed mostly unremarkably. Our best guess as to this result is that the high frequency of the noise makes many of the local extrema that TopoLines is trying to preserve unimportant for these data. 

Among the conventional smoothing methods, it is relevant to note that for the retrieve value task, Gaussian smoothing performed reasonably well overall, and for finding extrema, Douglas-Peucker performed well. Among the other methods, uniform subsampling, cutoff filter, and median filter, none performed consistently well at either task on multiple data types. We, therefore, recommend care in choosing to use them, at least for the tasks we evaluated.

\section{Conclusions}
\label{sec:conclusions}

In conclusion, we presented a topology-based line chart smoothing method called TopoLines. In the process, we showed that TopoLines has the potential to perform well for certain visual analysis tasks. However, all of these methods, including TopoLines, would benefit from an evaluation framework that considers a broader set of tasks and perceptual differences resulting from their use. In the future, we would like to build upon our current tasks list and run user studies to evaluate how the effect of smoothing on line charts are perceived. We hope to formulate a set of guidelines, based on these studies, that would be helpful for deciding which smoothing methods are best to use in practice.

\begin{comment}

\section{Discussion}

Based upon our analysis, TopoLines, low-pass filtering, and, to a slightly lesser extent, Douglas-Peucker all performed well. Even so, these methods have limitations, not necessarily accounted for in the evaluation. For example, low-pass filtering can develop ``ringing'' artifacts near the boundary, and determine the smoothing threshold in the frequency domain can be challenging. Nevertheless, the important relationship between all 3 approaches is that they tend to preserve large peaks in the data. Given this observation, TopoLines, whose primary objective is to optimize the preservation of high amplitude peaks, has an advantage. However, a deeper analysis of user preference is required to fully optimize the result, particularly with respect to the large design space available with smoothing, since the only requirement is that the output becomes monotonic.
%\fix{While our claim that TopoLines provides a more accurate representation for the smoothing of line charts, we believe a user study is both necessary and helpful in understanding how users interpret data at the intersection of scalar functions. In addition to providing a numerical analysis of TopoLines, we will also conduct a user study through Amazon's Mechanical Turk to measure users' preference for TopoLines when compared to other filtering techniques.}

\end{comment}

\section*{Acknowledgments} 
We would like to thank Bei Wang for providing valuable feedback on this project. This work was supported in part by a grant from National Science Foundation (IIS-1513616 and IIS-1845204).

\bibliographystyle{eg-alpha-doi}  
\bibliography{main}

\newpage

\end{document}